\documentclass[12pt,preprint]{aastex}

\newcommand{\eq}{\begin{equation}}
\newcommand{\feq}{\end{equation}}
\newcommand{\eqn}{\begin{eqnarray}}
\newcommand{\feqn}{\end{eqnarray}}

\shorttitle{Global Alfvenic heating of magnetic funnels}
\shortauthors{Elfimov, A.G., Galv\~ao, R.M.O., Jatenco-Pereira, V., Opher, R.}

\begin{document}

\title{ \bf Global Alfv\'{e}n Wave Heating of the Magnetosphere of
Young Stars}

\author{
A.G. Elfimov$^*$, R.M.O.Galv\~ao$^*$, V. Jatenco-Pereira$^+$, R.Opher$^+$}

\affil{$^+${\it Instituto de Astronomia, Geof\'{\i}sica e Ci\^encias Atmosf\'ericas}\\
{\it Universidade de S\~{a}o Paulo, 05508-900 SP, Brazil}\\
$^*${\it Instituto de F\'{\i}sica,
Universidade de S\~{a}o Paulo, 05315-970 SP, Brazil}}

\email{jatenco@astro.iag.usp.br}



\begin{abstract}

Excitation of a Global Alfv\'{e}n wave (GAW) is proposed as a viable mechanism 
to explain plasma heating in the magnetosphere of young stars. The wave and 
basic plasma parameters are compatible with the requirement that the 
dissipation length of GAWs  be comparable to the distance between the shocked 
region at the star's surface and the truncation region in the accretion disk. 
A two-fluid magnetohydrodynamic plasma model is used in the analysis. A 
current carrying filament along magnetic field lines acts as a waveguide for the 
GAW. The current in the filament is driven by plasma waves along 
the magnetic field lines and/or by plasma crossing magnetic field lines in the 
truncated region of the disk of the accreting plasma. The conversion of a 
small fraction of the  kinetic energy into GAW energy is sufficient to heat 
the plasma filament to observed temperatures. 

\end{abstract}


\section{Introduction}

\noindent In the present popular model for classical T Tauri stars (CTTSs),
the so called class II objects \cite{Lada87}, a central young star is
surrounded by a geometrically thin accretion disk. The disk is disrupted at a given
radius by the  magnetic field of the star. For smaller radii the accretion flow
follows the magnetic field lines of the star until it impacts onto the stellar
surface Ghosh \& Lamb (1979); K\"onigl (1991). This magnetospheric accretion
model explains observational signatures seen in some CTTSs as, for example,
the excess of optical and \- ultraviolet continuum flux (veiling), as well as redshift
absorption features in emission line profiles (Muzerolle $et\;al.$ 1998; Hartmann,
Hewett \& Calvet 1994). 
The plasma of the accretion disk
flows along the dipole magnetic field lines towards the magnetic
poles of the star. It is inhomogeneous in the plane perpendicular
to the magnetic field lines, but although somewhat more homogeneous along the field
lines, cavities and plasma condensations (plasmoids), aligned
by the magnetic field, may be present. These plasmoids are distributed 
randomly in the plane perpendicular to the plasma flow. Waves excited by the shock of the
plasma flux with the star's surface can propagate upstream and interact with
 the plasmoids, creating a current, which in turn, will create 
current - carrying filaments. An upstream or downstream electric current can
be driven by the waves or by the $\bf v\times B$-MHD effect in
the truncated region of the disk. Such filaments
have been observed in the laboratory (see, e.g.
Fadeev $et\;al.$ (1965); Rosenberg $et\;al.$ (2001)). The occurrence of filaments has also
been predicted in simulations of rotating magnetized disks (Machida $et\;al.$ 2000).

Hartmann, Hewett \& Calvet (1994) and, more recently, Muzerrole $et\;al.$ (1998) carried out
calculations on magnetosphere accretion models, solving the radiative transfer
equations in the Sobolev approximation. Theoretical Balmer
line profiles obtained were found to be in reasonable agreement with
observations. The major uncertainty in radiative magnetosphere model
calculations is the temperature profile along the tube. Martin (1996)
calculated the temperature profile from the energy balance of the gas, including
heating by adiabatic compression of magnetic field lines, Balmer
photoionization, and ambipolar diffusion. However, the values of the
temperature that he obtained were too low to explain the observed spectra. Thus,
additional heating mechanisms inside the tube are needed.

The line-averaged electron temperature in the magnetosphere taken from optical 
measurements, is about $0.8$ eV (see, e.g. Hartmann, Hewett \& Calvet 1994;
Martin 1996; Muzerrole $et\;al.$ 1998). Balancing the heating rate with the black body 
emission at a
temperature $\approx 0.8$ eV, we find that about 5$\%$ of the flux energy
is sufficient to heat the downstream flux to a few eV. The main
consumption of the flux energy is spent on plasma ionization. 
Using  laboratory experimental data for the
ionization - recombination processes (see Delcroix 1965), we can estimate
the energy necessary for the plasma ionization rate to be about $50\%$ in the
filament for electron temperatures of about $1.6$ eV. We use here a theoretical
temperature profile, varying from $\sim 7500$
K to $\sim 8300$ K, to reproduce the optical line features observed.

Taking into account the evidence for turbulence in the observations, as well as 
the necessity for an additional heating mechanism, Vasconcelos, Jatenco-Pereira \&
Opher (2002) suggested that Alfv\'en waves may be important in the heating of 
magnetic flux tubes of CTTSs. They studied the
possibility that the waves generated at the star's surface due to the
shock are produced by the accreting matter, as suggested by Scheurwater \& Kuijpers
(1988).

Various damping mechanisms for Alfv\'en waves have been suggested in the
literature, such as Alfv\'en resonant heating of solar loops (Hollweg 1990;
Elfimov $et\;al.$ 1996), wave damping by phase mixing (also in the solar context), and
cyclotron heating, occurring as an Alfv\'en wave travels down a magnetic field
gradient until its frequency matches that of the cyclotron resonance of helium or 
some other plasma specie due to the decreasing ion-cyclotron resonance
(magnetic beach). In addition to the more conventional collisional and
viscous-resistive Alfv\'en wave damping (Osterbrock 1961), Vasconcelos $et\;al.$ (2002)
concentrate on
nonlinear and turbulent damping (Vasconcelos, Jatenco-Pererira \& Opher 2000). 
Nonlinear and turbulent damping have also been studied by Jatenco-Pereira \&
Opher (1989a); Jatenco-Pereira, Opher \& Yamamoto (1994) 
in the solar wind, Jatenco-Pereira \& Opher (1989b) in protostellar winds, 
Jatenco-Pereira \& Opher (1989c) in late-type giant stars,
dos Santos, Jatenco-Pereira \& Opher (1993a; 1993b) in  Wolf-Rayet stars,
Gon\c calves, Jatenco-Pereira \& Opher (1993a; 1996) in quasar clouds and 
Gon\c calves $et\;al.$ (1993b)  in extragalactic jets .

The (GAW) has not been previously analyzed as a possible 
mechanism for the energy transport from the shock and truncation regions to the
magnetosphere of T Tauri stars. They were predicted in numerical calculations (Ross $et\;al.$
1982) 
using a cylindrical model for magnetized plasmas with an axial current and were 
observed later in tokamak experiments. GAWs are excited in plasmas 
in a helical magnetic field, with the poloidal component produced by an axial 
current, $B_{\theta,ef}\approx 2\pi aB_z/L$, where $B_z$ is the axial magnetic 
field component and $a$ and $L$ are the plasma radius and length, 
respectively. The so-called tokamak safety factor, 
$q=rB_{z}/RB_{\theta}$, is considered to be unity. In cold plasmas with 
temperatures less than $\sim $ 0.1 eV, GAWs dissipate via 
electron-ion collisions. However, in hot plasmas, when the Alfv\'{e}n velocity, 
$c_A=B/(4\pi m_in_e)^{1/2}$, is of the order of the electron thermal velocity, 
$v_{Te}=(T_{e}/m_e)^{1/2}$, GAWs dissipate via electron Landau damping. The 
electron Landau damping mechanism is very effective when waves interact with a 
group of electrons that have velocities approximately equal to the wave phase 
velocity, $\omega/k_{\parallel}= c_A$, where $\omega$ is the frequency and 
$k_{\parallel}$ is the wavenumber parallel to the magnetic field lines. 
Electromagnetic field components related to GAWs are  proportional to 
$\exp[i(m\theta +k_zz-\omega t)]$, with poloidal $m$ and axial $k_z$ wave 
numbers. The requirement for these waves to propagate is that $m=-1$ and 
$k_z<0$, in the case of current flowing along the magnetic 
field. A GAW has a very large quality factor, ratio of frequency to dissipation 
decrement, $Q=\omega/\gamma\gg 1$ and a very long dissipation length in comparison 
to their wave length, $L_A \gg 1/k_{\parallel}$.  Thus it is a very good candidate to 
explain plasma heating far from a 
source, such as in solar loops \cite{Hollweg, Elf96}. In addition a GAW 
is able to drive the current, which can be calculated by balancing 
the wave momentum transfer force with the electron-ion friction 
force (Elfimov $et\;al.$ 1996).

\section{Magnetosphere Accretion Model}

\noindent We study here a two-fluid magnetohydrodynamic  model of a young 
star, accreting plasma (see Fig. \ref{fig1}). The standard parameters of CTTSs, similar to 
those used by Vasconcelos $et\;al.$ (2002), are assumed: star radius 
$R_{\star}=10^{11}$ cm, 
truncated radius $R_{tr}=2 \times 10^{11}$ cm, truncated depth $\Delta_{tr} 
\approx 3\times 10^{10}$ cm, magnetic field at the star's surface $B_{st}=1$ kG  
and $B_{tr}\approx 0.1$ kG in the truncated region, where the electron 
density is ${\bar n}_{tr}=3.0\times {10}^{11}$ cm$^{-3}$. The length $L_{mf}$ of the 
magnetic field lines from the shock at the star's surface to the truncated 
region is $\approx 1.2 R_{\star}$. The velocity of the shock is $v_{sh}\approx 
2.5\times 10^7$ cm s$^{-1}$ at the star's surface and $v_{tr}\approx 3.5\times 
10^7$ cm s$^{-1}$ in the truncated region. The ionization rate parameter $<\sigma 
v>_{ion}$ for $T_e=1.5$ eV is $\sim 3\times 10^{-13}$ cm$^3$ s$^{-1}$ (Delcroix 1965). 
The accretion energy density is defined by the ratio of the 
energy flux density to the path length $L_{mf}$. Balancing ninety percent of 
the accretion power density $m_i n_0 v_{tr}^3/ L_{mf}$ with the ionization 
power density $<\sigma v>_{ion}n_e n_0 E_{ion}$, we obtain a plasma with an 
average density along the magnetic field of $\tilde n \approx 3.3\times 10^{11}$ cm$^{-3}$.



\placefigure{fig1}

We assume that the magnetic field connecting the truncated disk and the star's 
surface consists of filaments, along which the GAW propagates. The filament 
current can be driven by the wave ponderomotive force (see discussion below) 
and/or by a magnetodynamo in the truncated region of the disk. The current, 
$I_{[A]}=5aB_{\theta,ef}$, and current density, $j_{oh}=I/\pi a^2 \approx 
16$ statA cm$^2$, in a filament of radius $a$ can be produced by the electric 
field, $j/\sigma\approx 10^{-12}$ statV cm$^{-1}$, that is sufficient to create the 
required poloidal magnetic field $B_{\theta,ef}$ for the GAW excitation. This electric 
field can be generated by the induction of a perpendicular electric field, 
$E_{\perp}=B_{tr}v_{tr}/c \approx 0.05$ statV cm$^{-1}$, 
in the truncated region of the disk due to the magnetodynamo $\bf v\times B$ effect 
(see Fig. \ref{fig1}). A difference of potential on 
the scale of the filament radius is of the order of $2.5\times 10^3$ statV. 
This potential is much higher than is necessary and can generate a filament 
current with a hot electron tail and a characteristic velocity higher than that of the 
Alfv\'{e}n velocity $c_A$, as was found in the plasma 
focus device. We expect that the electrons in the tail will remain hot in the 
filament by their interaction with the GAW.

For the parameters of the star assumed above, the phase velocity of a GAW,
$c_A(R_{tr})$, is $\sim 3\times10^{8}$ cm s$^{-1}$ at the center of the filament. Taking
into account the plasma resistance (see, e.g. Ginsbrug 1961), the
dissipation length of the Alfv\'{e}n waves can be estimated to be
\begin{equation} L_A\approx
8\pi\frac{\sigma c_A}{k_{\parallel}^2c^2}\approx 1.2\times 10^{11} \; {\rm cm},
\end{equation}
where
$$\sigma=\frac{\omega_{pe}^2}{4\pi\nu_{ei}}\approx 1.6\times 10^{13}
T_e^{3/2} \; {\rm s}^{-1},\,\,\ k_{\parallel}=3\times 10^{-5} \; {\rm cm}^{-1} .$$
This $k_{\parallel}$ is a maximum value, determined by the condition that the
dissipation length be sufficient  so as to explain the plasma heating along the
magnetic field from the truncated region to the star's surface.

Assuming a conversion of 5$\%$ of the total accreting plasma energy flux, 
$S=(1/2)n_n m_{i} (v_{sh}^3 +v_{tr}^3)$, into the Alfv\'{e}n wave flux, 
$S_A=P_A\,L_A$, we obtain a wave dissipation rate of $P_A=0.0039$ 
erg cm$^{-3}$ s$^{-1}$. This dissipation rate produces a wave momentum 
transfer force $P_Ak_{\parallel}/\omega$ (see Elfimov $et\;al.$ 1996). Balancing 
the wave momentum transfer force with the electron-ion friction force 
$\nu_{ei}m_en_e V_e$, we obtain the density of the driven current,
\begin{equation}
<j_{cd}> = - \frac{ |e| k_{\parallel} P_A}{m_e \nu_{ei,ef}\omega} \; .
\end{equation}
Using a reduced friction, $\nu_{ei,ef}\approx \nu_{ei}(v_{Te}/c_A)^{3/2}$, for 
the electron tail with effective velocity  $\approx c_A$ and
$\omega=k_{\parallel}c_A =9\times10^3$ s$^{-1}$  ($c_A= 3. \times 10^8 $ cm s${-1}$),  we
obtain the required current density, $<j_{cd}>\approx 16$ statA cm$^{-2}$, in the
filament. In this case, the electron tail that transports the current
along the filament can be supported by the GAW via the electron Landau
damping, while the current in the filament can be driven by the GAW.

Here, we also present the results of calculations for the GAW in the filament. 
In order to calculate the wave field and dissipation, we use 
the eikonal model to analyze the wave propagation along the filament, 
exp$[i(m\theta+\int k_z dz-\omega t)]$, where $z$ is the coordinate along the 
filament axis and $\theta$ is the poloidal angle. In this case, in order to simplify 
the GAW calculations using numerical codes \cite{Ges,Gal}, instead of a dipole 
magnetic field, we assume a cylindrical model for the filament plasma with the 
current flowing along a magnetic field that has a helical magnetic field line 
configuration.  To calculate the wave characteristics along the 
filament, we take the major radius $R_{0}=10^{11}$ cm comparable to the star's
radius and the minor radius $a=5\times 10^5$ cm. The calculations 
are carried out with the electron density profile of the form 
$n_{e}=n_{0}(1-r^{2}/a^{2})+n_{a},$ with $n_{0}=5\times {10}^{11}$ cm$^{-3}$ and 
$n_{a}=5\times {10}^{10}$ cm$^{-3}$. The temperatures are taken as 
$T_{e0}=1.6$ eV and $T_{i0}=0.5$ eV. The 
ion density is chosen so as to satisfy the requirement of charge neutrality, 
$Zn_{Z}+n_{i}=n_{e}$, where $n_{i}$ and $n_{Z}$ are the densities of 
hydrogen and the other ions, respectively. In accordance with the Spitzer 
conductivity $\propto T_e^{3/2}$, the current density profile has the form 
$j=j_{0}(1- r^{2}/a^{2})^{3}$. The value of the safety factor, 
$q=rB_{z}/RB_{\theta}$, is assumed to be unity at the filament axis and  to 
monotonically increase up to four at the plasma boundary. In Fig. \ref{fig2}, we show the 
characteristic profile of the magnetic field as well as the dissipated power of the GAW for 
the filament parameters discussed above.



\placefigure{fig2}

In Fig. \ref{fig2}, we observe the peaking of the dissipation at the filament center, which
predicts filament center peaking of the
electron temperature distribution, as well. The GAW heats the core of the filament to 
higher temperature, while reducing the plasma temperature 
at the filament border due to heat diffusion. The axial wavenumber of the GAW, 
$k_z=\omega/c_A$, varies slowly along the magnetic field lines since it
depends on changes in 
the local magnetic field and density.

\section{Conclusions}
Our analysis of GAWs in the magnetospheres of CTTSs shows
that:\\
a) The energy of the shock (and/or accretion disk region) can be transported 
from the surface of the star (and/or accretion disk region) to the 
magnetosphere region by GAWs via current carrying filaments;\\
b) The dissipation length of GAWs is on the order of the truncation
radius;\\
c) The conversion of only a few percent of the shock energy into GAWs is required
in order to heat the filaments up to the observed temperature of the star's magnetosphere;\\
d) The current in the filament may be induced by the electric field, created
by the $\bf v\times B$-magnetodynamo effect in
the truncated region of the disk. However, the major part of the current in
the filaments is driven by GAWs.

\acknowledgments The authors thank the Brazilian agency FAPESP 
(Proc. No. 00/06770-2) for support. VJP and RO thank the Brazilian 
agency CNPq and the project PRONEX (41.96.0908.00) for partial
support.

\newpage
\figcaption[f1.ps]{Sketch of a young accreting star of radius
$R_\star$ and truncation depth $\Delta_{tr}$. The electric field, designated by $\oplus$,
is generated by the $\bf v\times B$-effect due to the interaction of the magnetic
field lines from the star with the accreting plasma in the truncated region of the
disk. \label{fig1}}

\figcaption[f2.ps]{Plot of $B_{r,\theta}$-components of the GAW field and 
dissipated power across the filament cross-section for a frequency of 1.4 kHz 
and the parameters $B= 1$kG, $q_0=1.0, n_{0}=5\times {10}^{11}$ cm$^{-3}$,
$T_{e0} = 1.6$ eV, and $T_{i0} = 0.5$ eV. \label{fig2}}

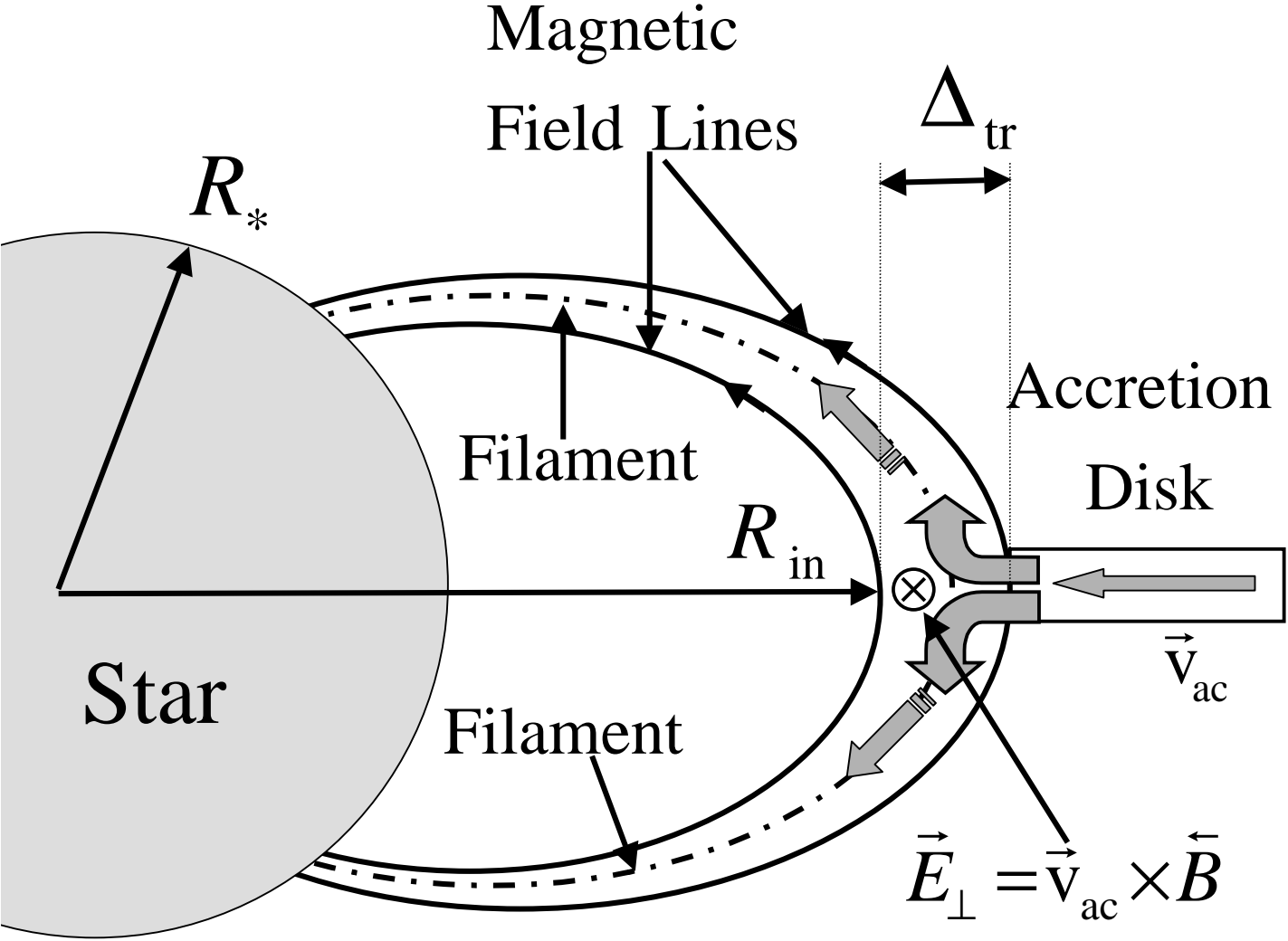
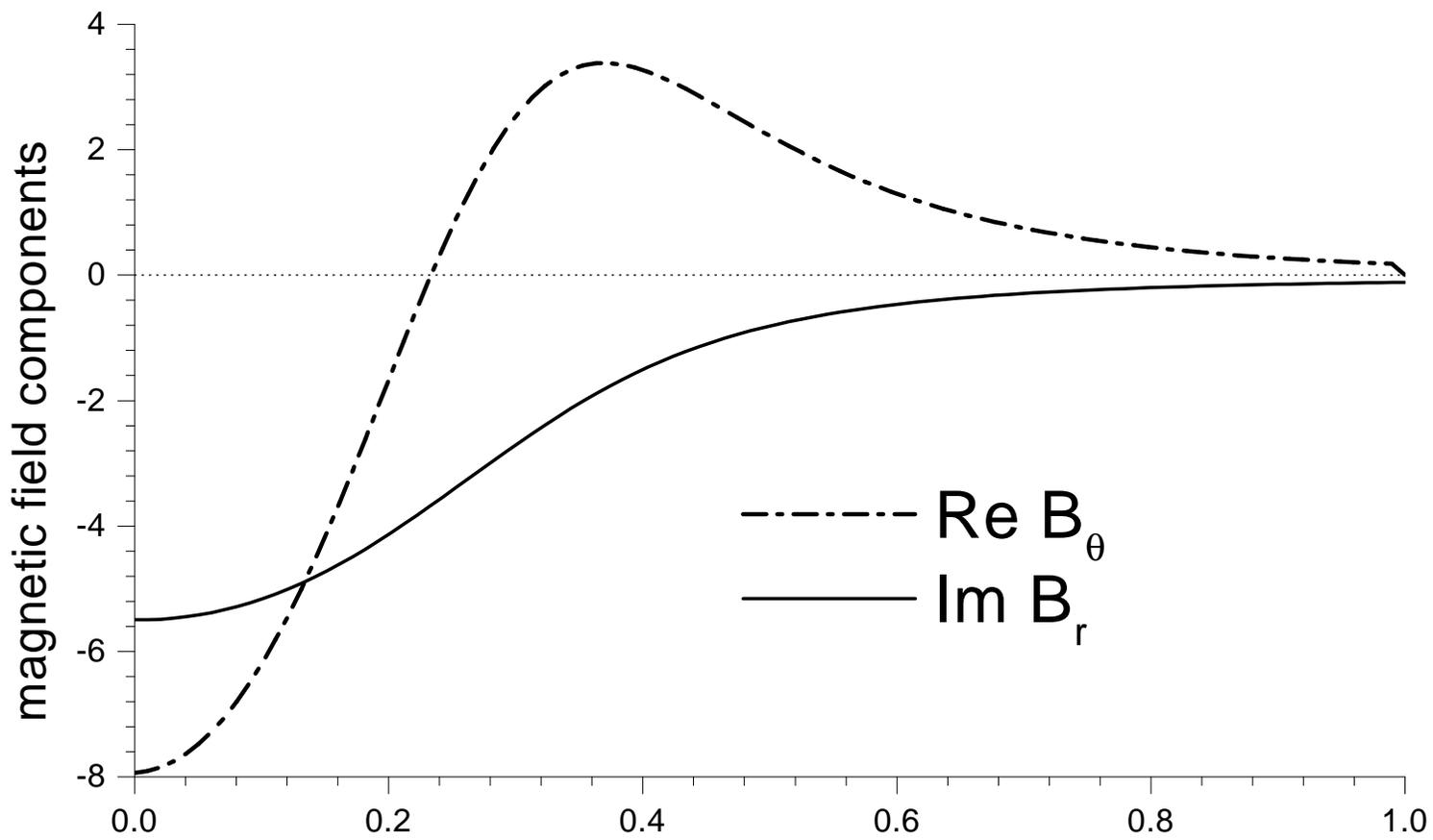
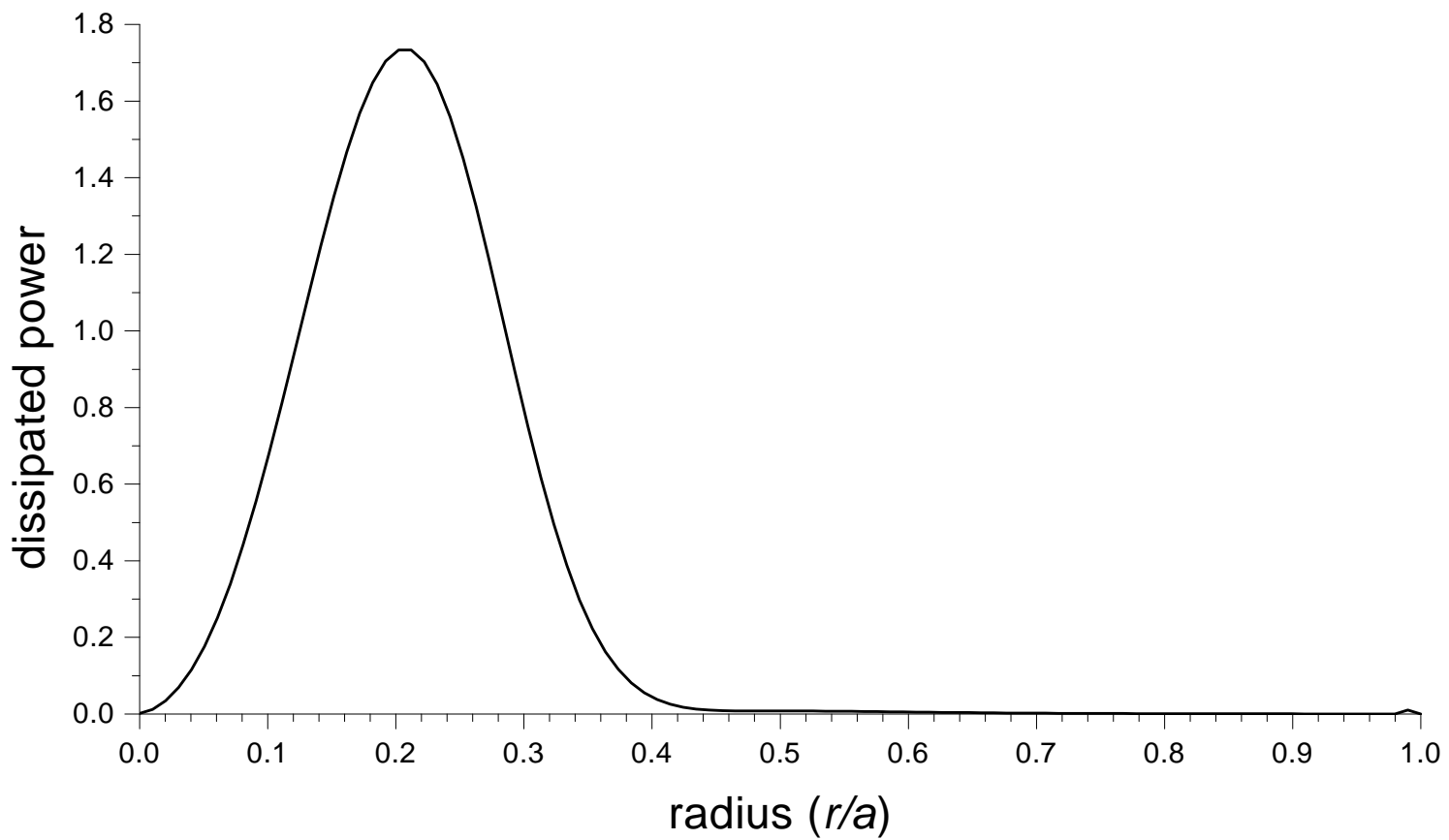
\end{document}